\documentclass{article}
\usepackage[preprint]{spconf}
\usepackage{amsmath,graphicx, multirow}
\usepackage{atbegshi,picture}

\newcommand{\logVal}{L}
\newcommand{\entropyVal}{E}
\newcommand{\numSig}{S}
\newcommand{\subblock}{\mathcal{S}}
\newcommand{\block}{\mathcal{B}}
\newcommand{\REst}{{R_\mathrm{est}}}

\title{A Low-Parametric Model for Bit-Rate Estimation of VVC Residual Coding}
\name{Fabian Brand, Christian Herglotz, Andr\'e Kaup}
\address{Multimedia Communications and Signal Processing, Friedrich-Alexander-Universit\"at Erlangen-N\"urnberg}
\AtBeginShipout{\AtBeginShipoutUpperLeft{%
		\put(\dimexpr\paperwidth-20cm\relax,-1.5cm){\parbox[t]{19cm}{\copyright 2022 IEEE.  Personal use of this material is permitted.  Permission from IEEE must be obtained for all other uses, in any current or future media, including reprinting/republishing this material for advertising or promotional purposes, creating new collective works, for resale or redistribution to servers or lists, or reuse of any copyrighted component of this work in other works.}}%
}}

\begin{document}

\maketitle
\begin{abstract}
There are many tasks within video compression which require fast bit rate estimation. As an example, rate-control algorithms are only feasible because it is possible to estimate the required bit rate without needing to encode the entire block. With residual coding technology becoming more and more sophisticated, the corresponding bit rate models require more advanced features. In this work, we propose a set of four features together with a linear model, which is able to estimate the rate of arbitrary residual blocks which were compressed using the VVC standard. Our method outperforms other methods which were used for the same task both in terms of mean absolute error and mean relative error. Our model deviates by less than 4 bit on average over a large dataset of natural images.
\end{abstract}
\vspace{-0.5cm}
\section{Introduction}
\vspace{-0.2cm}
Bit rate estimation is a common problem in image and video compression. There are multiple tasks in a video coder where knowledge of the resulting bit rate is beneficial. This includes for example rate-control~\cite{WiegandSJ2003_Rateconstrainedcoder}, where the encoder has to match a specified rate at the output. Here, the encoder has to know the resulting rate of a coding decision in order to accurately choose the coding parameters for optimal rate-distortion behavior, a process which is known as rate-distortion-optimization~\cite{SullivanW1998_Ratedistortionoptimization,LiOH2009_LaplaceDistributionBased}. One possibility to obtain the rate needed for a certain decision is to perform the compression until the end and measure the rate directly. While this is certainly the most accurate method, following the coder to the end is often time-consuming, when performed too often. In such cases, a robust, simple and accurate rate model is useful to estimate the required rate. 

One example where the rate has to be estimated multiple times is rate-distortion optimized quantization (RDOQ)~\cite{RamchandranV1994_Ratedistortionoptimal,StankowskiKD2015_Ratedistortionoptimized,SchwarzNM2019_ImprovedQuantizationTransform}. Here the quantized values themselves are subject to rate-distortion optimization. For each tested configuration, both rate and distortion have to be evaluated. Another application targets video formats with irrelevant content. For example, some projection formats in 360$^{\circ}$ video coding map the content to a 2D plane, where parts of the sequence are irrelevant for the reconstruction of the 360$^{\circ}$ content \cite{HerglotzJC2019_EfficientCoding360}. Similar regions can be found in videos generated for video-based point cloud compression \cite{LiLL2020_Occupancymapbased}. In such sequences, several blocks may contain both relevant as well as irrelevant pixels, where the exact pixel values in the irrelevant region are of no importance. Depending on the number of irrelevant pixels, this can lead to a manifold of optimal representations in the DCT domain, where the best representation must be determined for optimal rate savings. Using a simple rate model as proposed in this paper instead of performing the arithmetic encoding chain for all DCT representations will decrease the encoding time significantly.

Coding of frequency coefficients has become more sophisticated with ongoing standard development. While JPEG only uses run-level coding to encode the values~\cite{Wallace1992_JPEGstillpicture}, VVC uses multiple coding passes and can therefore exploit the sparsity of images in the frequency domain, in particular in high frequency areas. Additionally, VVC uses context adaptive binary arithmetic coding (CABAC)~\cite{MarpeSW2003_ContextBasedAdaptive}, which is able to take the previously coded signals into account to improve compression efficiency. So in the end, the required rate for one residual block follows more complicated relationships and is not even deterministic, since  different contexts in CABAC can yield different rates for the same residual signal.

In this work, we limit ourselves to models which estimate the rate needed to transmit a transformed and quantized residual block in VVC. Our goal is to design a rate model, which only uses the quantized frequency coefficients and which is valid for all block-sizes. In the following, we first introduce our model and the features on which it is based. We then compare our model against other methods performing a similar task and perform ablation studies demonstrating the impact of the individual features. In the end, we show the versatility and robustness of our model in various experiments.
\vspace{-0.2cm}
\section{Related Work}
\vspace{-0.2cm}
Rate models were often examined in the context of rate-control. This field was dominated for a long time by $\rho$-domain models \cite{HeKM2001_Lowdelayrate,LiuGL2010_Lowcomplexityrate}. Here, a linear relationship between the number of non-zero frequency coefficients and the rate is assumed. In these models, the proportionality factor is often not constant for all contents but rather determined dynamically based on various image and video properties, like texture or previously coded pictures \cite{LiuGL2010_Lowcomplexityrate}. 

Another class of rate-control models are the so-called $\lambda$-domain models \cite{LiLL2014_Lambda,Sanchez2018_RateControlHevc}. These models assume a usually exponential relationship between the rate and the Lagrangian factor $\lambda$ of the rate-distortion optimization. As for the $\rho$-domain model, the parameters of the model are often estimated using various image properties.

Recently, learning-based methods for the use in video coding have been proposed. These methods often tackle prediction problems, such as \cite{BrandSK2020_IntraFrameCoding} which proposes a deep-learning-based intra-prediction method for use in VVC. Also, VVC includes Matrix Intra Prediction (MIP) \cite{PfaffSS2019_CE3Affinelinear}, which includes a trained matrix. These methods generate prediction signals which are optimal if the residual can be compressed well. These methods therefore require a loss function which accurately estimates the rate of a residual block. Here, it is desirable to have fixed model parameters unlike the $\rho$- and $\lambda$-domain models, that do not depend on other image characteristics. One such model was proposed in \cite{HellePS2019_IntraPicturePrediction} which was also used to train MIP.

\section{Rate Estimation Model}
\vspace{-0.2cm}
The rate estimation model we propose in this work consists of four features and a bias, which form a linear model with five parameters. We chose a linear model due to its low complexity and simplicity in training. With only five parameters, we expect the model to be robust against overfitting and to generalize well. The features are hand-crafted and inspired by the process of residual coding in VTM. The features are all computed on sub-block level, meaning, each block is divided into $4\times4$ sub-blocks before feature computation. The feature of the block $\block$ itself is then the sum of the features of all sub-blocks. This design concept grants us the possibility to find a model which is valid for all block-sizes. In the following, the set of all coefficients $c$ in a sub-block is denoted as $\subblock$.  

As first feature we choose to use the number of non-zero coefficients after quantization. This feature is similar to the parameter of the $\rho$-domain model. The motivation of the feature is that in the beginning of the compression scheme, non-zero coefficients are signaled and all subsequent bits are only required for there so-called significant coefficients. This has been proven a good feature in the past, in particular in the $\rho$-domain model. In the following, this feature is denoted as~$\numSig$. We can express this feature as
\begin{equation}
\numSig = \sum_{\subblock\in \block}\sum_{c\in\subblock}~\begin{cases}
~~0 & \mathrm{if~} c \neq 0\\
~~1 & \mathrm{else}
\end{cases}.
\end{equation}

As second feature we use the sum of the binary logarithm of all significant coefficients. This feature estimates how much rate has to be spent to transmit each value. We denote this feature as $\logVal$ with:

\begin{equation}
\logVal = \sum_{\subblock\in \block}\sum_{c\in\subblock}\max(0,\log_2|c|).
\end{equation}

The next two features describe the positions and distributions of the coefficients on sub-block-level. For each sub-block, we perform a zig-zag scan and look at the last coefficient which is greater than zero. We then use the position of that coefficient as sub-block feature and sum up all the positions for the block feature. We denote this third feature as $Z$.

The fourth feature measures the percentage of coefficients in the sub-block which are strictly greater than 1 and computes the binary entropy function of the percentage:
\begin{equation}
	\entropyVal = \sum_{\subblock\in \block}H_2\left(\frac{C_1(\subblock)}{16}\right),
\end{equation}
where $H_2(p)=-p\log_2(p)-(1\!-\!p)\log_2(1\!-\!p)$ is the binary entropy function and $C_n(\subblock)$ is a function counting all coefficients with values greater than $n$ in the sub-block $\subblock$. In practice, this can easily be realized with a lookup table. 

Altogether, we use the four features in a linear model for rate estimation:

\begin{equation}
	\REst = a\cdot\numSig + b\cdot\logVal + c\cdot Z + d\cdot\entropyVal + e,
\end{equation}
with trainable parameters $a$, $b$, $c$, $d$, and $e$, the latter of which is a global offset.

\section{Experiments}
\vspace{-0.3cm}
\begin{table*}
	\centering
	\begin{tabular}{ll|cccc|cccc|cccc}
		&&\multicolumn{4}{|c}{$\rho$ domain \cite{HeKM2001_Lowdelayrate}}&\multicolumn{4}{|c}{Logistic \cite{HellePS2019_IntraPicturePrediction}} &\multicolumn{4}{|c}{Sub-Block Model (Ours)}\\
		QP$_\mathrm{t}$&QP$_\mathrm{e}$ & $P$ & MAE & MRE & time & $P$ & MAE & MRE & time & $P$ & MAE & MRE & time \\\hline
		22&22 & 0.9752&12.3&17.1\%&0.003& 0.9959&6.91&12.7\%&0.002& 0.9978 & 4.78 & 9.1\% & 0.009\\
		27&27 & 0.9845&8.0 &17.2\%&0.003& 0.9936&6.10&15.2\%&0.003& 0.9970 & 4.02 & 10.1\% &0.009\\
		32&32 & 0.9898&5.96&17.5\%&0.003& 0.9941&4.12&15.7\%&0.003& 0.9970 & 3.45 & 12.7\% &0.009\\
		37&37 & 0.9874&4.55&17.0\%&0.003& 0.9902&5.33&16.5\%&0.004& 0.9954 & 2.98 & 14.3\% &0.009 \\\hline
		\multirow{2}{10pt}{22}&27 &0.9830&8.42&17.7\%&- &0.9947&5.48&14.6\%&- &0.9967&4.07&10.9\%&- \\
		&32 &0.9905&6.70&18.8\%&-  &0.9928&4.01&16.1\%&- &0.9967&3.52&12.5\%&- \\
		\multirow{2}{10pt}{37}&27 &0.9849&8.9&18.1\%&- &0.9898&7.76&16.1\%&- &0.9968&4.60&11.6\%&- \\
		&32 &0.9904&6.01&17.9\%&- &0.9905&5.84&16.1\%&- &0.9969&3.57&12.3\%&- \\
	\end{tabular}
	\vspace{-0.2cm}
	\caption{Results of the rate estimation experiments. QP$_\mathrm{t}$ denotes the QP which was used for training, QP$_\mathrm{e}$ denotes the QP which was used for evaluation. The time is given in ms per block.}
	\vspace{-0.2cm}
	\label{Tab:Results}
\end{table*}
\subsection{Setup}
To evaluate our model, we compare it with two different models which are both estimating the rate based on transformed blocks. First, we use the $\rho$-domain model from \cite{HeKM2001_Lowdelayrate}. The $\rho$-domain model assumes a linear relationship between the percentage of non-zero quantized coefficients $(1-\rho)$ and the rate. Since this model was proposed for usage in MPEG-4, where all transform blocks were of size $4\times4$ and therefore constant, the model does not take the size of the block into account. We instead use the absolute number of non-zero coefficients as a feature. 

Also, we use the model which was suggested in \cite{HellePS2019_IntraPicturePrediction} as loss function to train intra prediction networks for MIP \cite{PfaffSS2019_CE3Affinelinear}. Since this model was initially proposed as a loss function for a neural network, a linear relationship to the actual rate was sufficient. In our case, however, we require an exact estimate. We therefore extended the model by two parameters, such that the rate for one block is now estimated by
\begin{equation}
	\REst = \sum_{(m,n)}\alpha\left|c_{m,n}\right|+\beta g\left(\gamma\left|c_{m,n}\right|+\delta\right)+\varepsilon,
\end{equation}
with the logistic function $g(\cdot)$. We therefore call this model the logistic model. At this point we want to note that this model, was designed as differentiable model and therefore does not use features based on thresholding, counting, or positions.

For our experiments we encoded 30 picture from the DIV2K dataset \cite{AgustssonT2017_NTIRE2017Challenge} using VTM 10.0 \cite{ChenYK2020_AlgorithmdescriptionVersatile} and QPs of 22, 27, 32 and 37. This yields between 450,000 and 1,100,000 blocks per QP. We then perform a 5-fold cross validation and average the metrics. We train each model to minimize the mean squared error (MSE). For the $\rho$-domain model and our proposed method, we can use the pseudo-inverse for estimation, while the non-linear logistic model is trained using gradient descent. 

To evaluate the results, we use the Pearson correlation coefficient
\begin{equation}
P = \frac{\sigma_{R,\REst}}{\sigma_R\sigma_\REst},
\end{equation}
where $\sigma_{R,\REst}$ denotes the covariance of $R$ and $\REst$. We furthermore evaluate the mean absolute error MAE 
\begin{equation}
\mathrm{MAE} = \frac{1}{N}\sum_{i=1}^{N}\left|R_i-\REst_{,i}\right|
\end{equation}
and the mean relative error MRE.
\begin{equation}
\mathrm{MRE} = \frac{1}{N}\sum_{i=1}^{N}\frac{\left|R_i-\REst_{,i}\right|}{R_i},
\end{equation}
where $R_i$ and $\REst_{,i}$ denote the $i\textsuperscript{th}$ sample in the test set and $N$ is the total number of tested blocks.

\subsection{Rate Estimation}
\begin{figure}
	\centering
	\includegraphics[width=0.8\columnwidth]{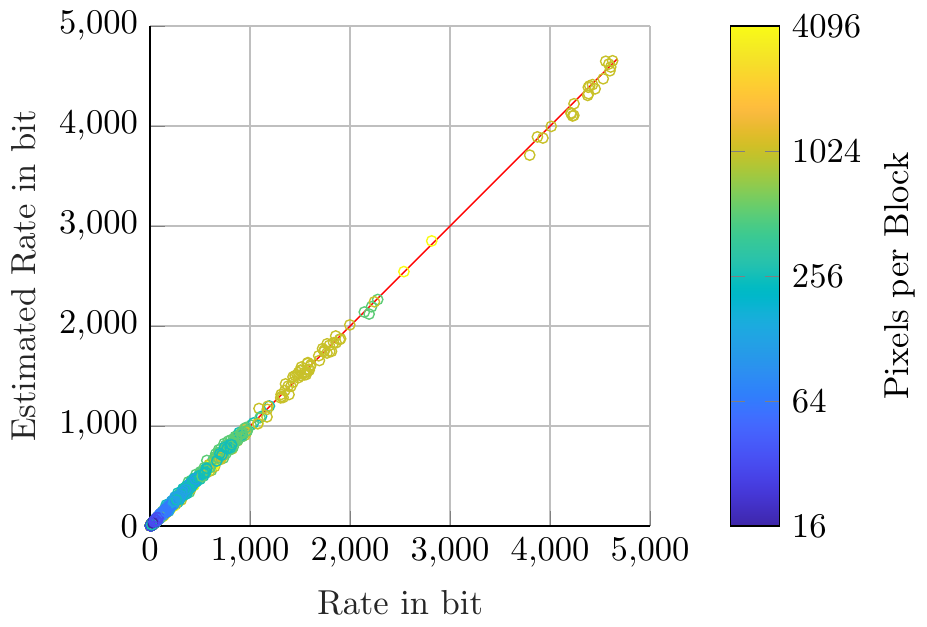}
	\vspace{-0.5cm}
	\caption{Estimation curve for our proposed rate model.}
	\vspace{-0.5cm}
	\label{Fig:Results}
\end{figure}
\begin{table}
	\vspace{-0.5cm}
	\centering
	\begin{tabular}{lccc}
		Feature Set & $P$ & MAE & MRE \\
		$\numSig$ & 0.9791 & 12.34 & 17.2\%\\
		$\logVal$ & 0.8995 & 33.03 & 48.36\%\\
		$Z$  & 0.9242 & 20.69 & 27.1\%\\
		$\entropyVal$ & 0.9361 & 25.02 & 45.8\%\\\hline
		$\numSig+\logVal$ & 0.9976 & 5.70 & 11.8\%\\
		$\numSig+Z$ & 0.9820 & 12.38 & 18.8\%\\
		$\numSig+\entropyVal$ & 0.9844 & 10.32 & 16.7\%\\
		$\logVal + Z$ & 0.9934 & 7.67 &  14.9\%\\
		$\logVal + \entropyVal$ & 0.95 & 19.92 & 36.8\%\\
		$Z+\entropyVal$ & 0.9710 & 15.11 & 26.7\%\\\hline
		$\numSig + \logVal + Z$ & 0.9977 & 5.3 & 11.2\%\\
		$\numSig + \logVal + \entropyVal$ & 0.9975 & 5.34 & 10.9\%\\
		$\numSig + Z + \entropyVal$ & 0.9878 & 10.40 & 16.6\%\\
		$Z + \logVal + \entropyVal$ & 0.9948 & 6.76 & 13.6\%\\\hline
		$\numSig + Z + \logVal + \entropyVal$ & 0.9978 & 4.93 & 10.5\%
	\end{tabular}
	\vspace{-0.3cm}
	\caption{Results of the ablation studies.}
	\vspace{-0.3cm}
	\label{Tab:Abl}
\end{table}

\begin{figure*}[t!]
	\centering
	\begin{tabular}{ccc}
		\includegraphics[width=0.28\linewidth]{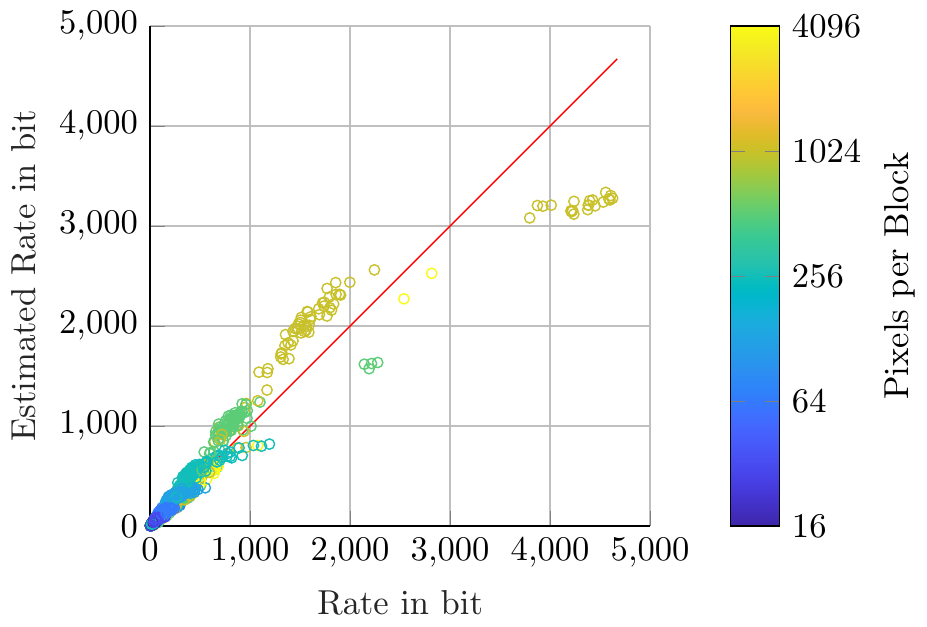} & \includegraphics[width=0.28\linewidth]{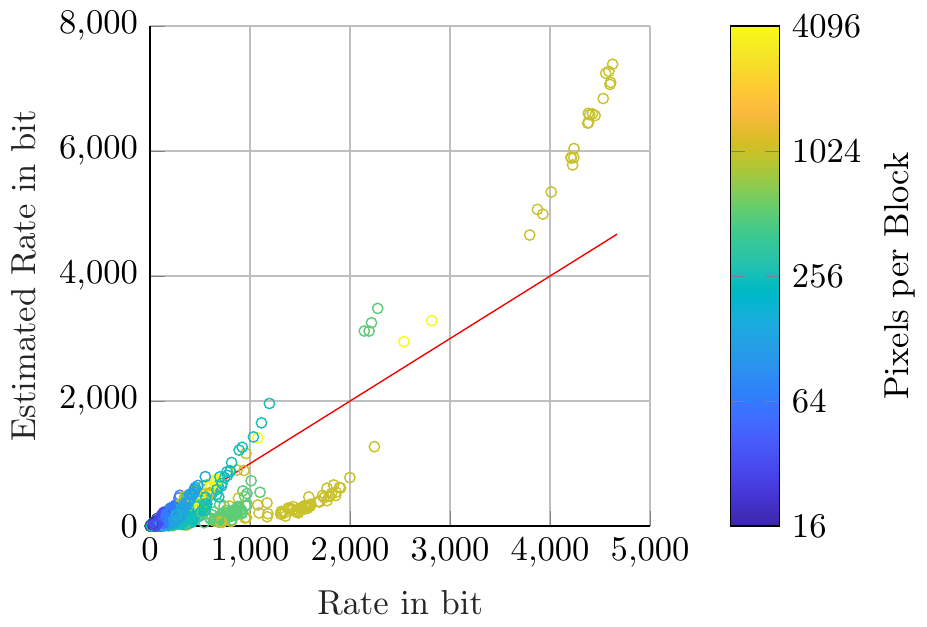} & \includegraphics[width=0.28\linewidth]{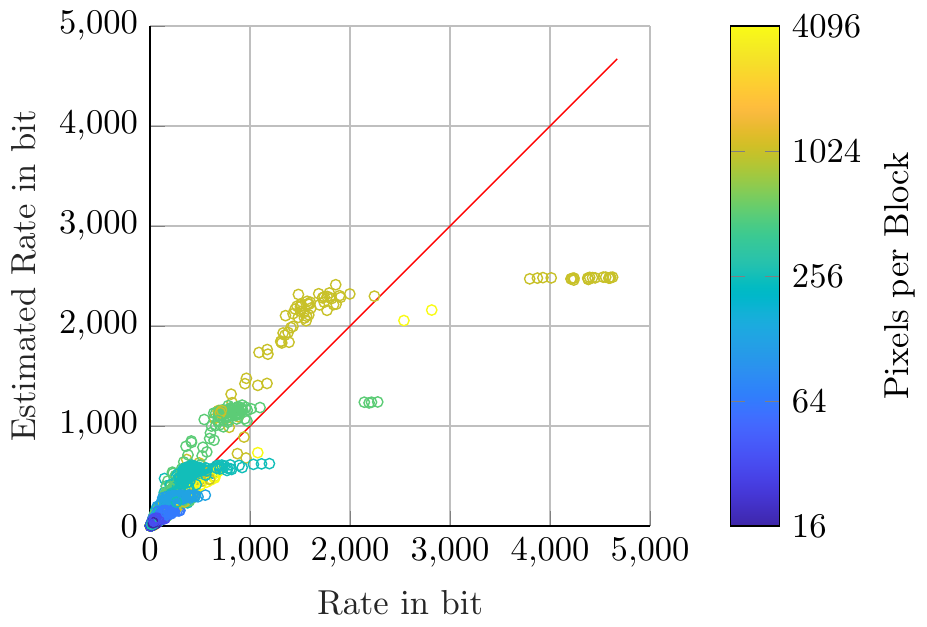} \\
		(a) $\numSig$ & (b) $\logVal$ & (c) $Z$\\ \includegraphics[width=0.28\linewidth]{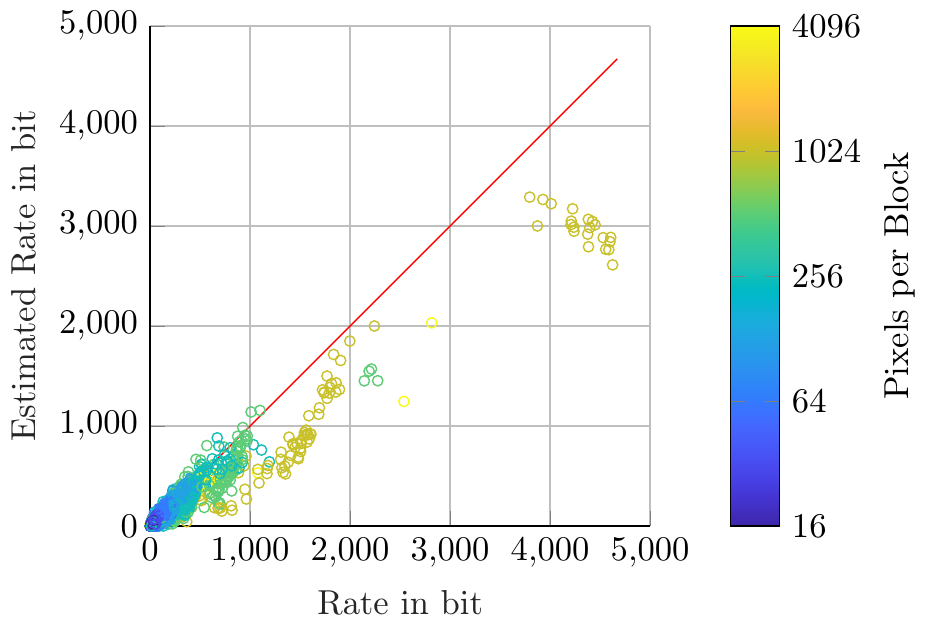} & \includegraphics[width=0.28\linewidth]{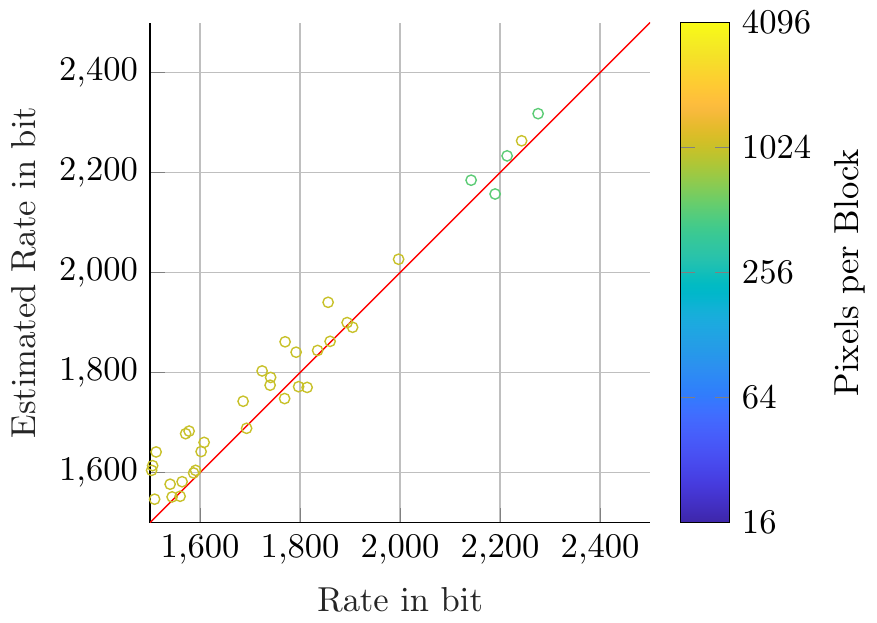} & \includegraphics[width=0.28\linewidth]{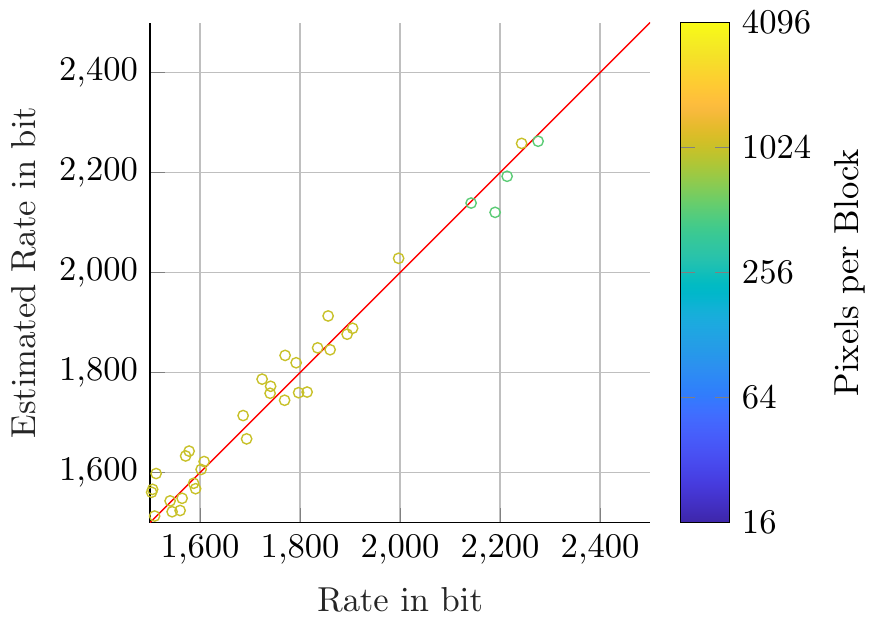}\\
		(d) $\entropyVal$ & (e) $\numSig + \logVal + Z$ (zoomed)& $\numSig + Z + \logVal + \entropyVal$ (zoomed)\\
		
	\end{tabular}
\vspace{-0.2cm}
	\caption{Estimation curves for different feature combinations.}
	\vspace{-0.4cm}
	\label{Fig:Abl}
\end{figure*}
In a first experiment, we compare the three models for each QP separately. In the 5-fold cross validation, both training set and test set were compressed with the same QP. We show the results evaluated on the test sets in Tab. \ref{Tab:Results}.

From this result we see that the $\rho$-domain model, even though it is a very simple model, still performs relatively well. This indicates that the number of non-zero coefficients is a good indicator and also works well if the parameters are trained on general images and not on specific blocks. 

The sigmoid model, which is a more complex model and takes the magnitude of the coefficients into account, performs better with a mean relative error of 12.1\% for QP 22 and 19.5\% for QP 37. The rise of the MRE with the QP indicates that the model does not perform well for small rates, since the rate average rate decreases with QP. 

Our proposed sub-block model performs better than the other two models in all metrics. Throughout all QPs, we achieve a correlation coefficient of $R>0.995$ and the relative error---even though it is also increasing with the QP---is between 2 and 5 percentage points below the MRE of the sigmoid model. Fig. \ref{Fig:Results} shows the measured bit rate against the estimated rate. The points are color coded according to the total number of pixels of that block. We see that our model is able to produce accurate results over all rates. The red line indicates the case of perfect estimation.

To compare the complexity of our methods, we conducted time measurements in our MATLAB implementation. Here, we see that our method takes about 3 times longer than the others. Note, that the runtime may vary in different implementations. The measurements show that our method is in a similar range than the others with several $\mu$s per block. 

To demonstrate the versatility of our model, we additionally perform a cross QP test. For this test, we train the models on images which were coded with QP$_\mathrm{t}\in\left\{22,37\right\}$ and evaluate on images which were coded with QP$_\mathrm{e}\in\left\{27,32\right\}$. The results show that all three methods are relatively robust to the QP of the training data. In most cases, we can observe a small degradation when we train with deviating QPs, however, this is always less than 2\% in MRE. On the other hand, we also see a few cases where the results are better. Slight deviations are expected since data may randomly produce a better fit. As expected, however, on average the quality degrades with a QP mismatch. In the end, our proposed method still outperforms the other methods in all metrics also in this case.
\vspace{-0.2cm}
\subsection{Ablation Study}
\vspace{-0.2cm}
Additionally, we perform ablation studies to demonstrate the descriptive power of each feature. The following experiments were performed with a QP of 22. The results in Tab. \ref{Tab:Abl} clearly show that the individual features are not well suited to estimate the rate. The best individual feature is the number of non-zero coefficients $\numSig$, which is equivalent to the $\rho$-domain model. In the table, we can see that especially $\logVal$ and $\entropyVal$ perform poorly by themselves. In Fig. \ref{Fig:Abl}, we show the results of a selection of cases to further illustrate the results. 

We can see that $\numSig$ and $Z$ severely underestimate the rate for large rates and that the behavior is non-linear. This is due to the fact that both features do not take the actual values of the coefficients into account. On the other hand, the model using only $\logVal$ leads to an overestimation for large rates. In Tab. \ref{Tab:Abl}, we see that $\numSig+\logVal$ and $Z+\logVal$ both give reasonably good results as both effects cancel out and form a better linear model. In Fig. \ref{Fig:Abl} (e) and (f), we see show the effect of the entropy based feature. This feature mainly influence the region for medium bit rates. We can see here that in these ranges, $\numSig + \logVal + Z$ leads to a slight overestimation, which can be fixed by taking the entropy-based feature into account. Note that all ablation study experiments were performed without a bias term. This also shows the importance of being able to add a constant term, as it is responsible for the improvement from 10.5\% to 9.1\% relative error.
\vspace{-0.1cm}
\section{Conclusion}
\vspace{-0.2cm}
In this paper, we proposed a novel set of features to estimate the rate required to compress a transformed residual block in VVC. Our model has only five parameters and can be trained easily and stably due to its low complexity and linearity. Other than in known approaches like $\lambda$-domain models, we can find fixed parameters to model the rate for all blocks, independent of the content.

In future work, the model accuracy could further be increased by taking the CABAC context into account and including it in the model. Since the context can change the required rate, no model which does not include the context can exceed a certain accuracy. Furthermore, the model can be extended to other video compression standards and the effect of our model in several scenarios, such as RDO, RDOQ or irrelevant region coding, both in compression performance and in runtime can be evaluated. The model we presented in this paper showed superior performance to other models, while remaining simple and linear. Therefore, a speedup and good performance can be expected in various applications.


\clearpage

\bibliographystyle{IEEEbib}

\end{document}